\documentclass[twocolumn]{article}

\usepackage{multirow}

\usepackage{amsmath}

\begin{document}

\date{November 10, 2017}

\title{Discrete Scaling Based on Operator Theory}

\author{Aykut~Ko\c{c},\thanks{Aykut Ko\c{c} (Corresponding Author) is with ASELSAN Research Center, Ankara, Turkey, e-mail: aykutkoc@aselsan.com.tr.}
$\:$ $\:$       Burak~Bartan,\thanks{Burak Bartan is with Electrical Engineering Department, Stanford University, Stanford, California}
        $\:$ $\:$   ~Haldun~M.~Ozaktas,
\thanks{Haldun M. Ozaktas is with Electrical Engineering Department, Bilkent University, TR-06800 Bilkent, Ankara, Turkey}}

\maketitle

\begin{abstract}
Signal scaling is a fundamental operation of practical importance in which a signal is enlarged or shrunk in the coordinate direction(s). Scaling or magnification is not trivial for signals of a discrete variable since the signal values may not fall onto the discrete coordinate points. One approach is to consider the discretely-spaced values as the samples of a signal of a real variable, find that signal by interpolation, scale it, and then re-sample. However, this approach comes with complications of interpretation. We review a previously proposed alternative and more elegant approach, and then propose a new approach based on hyperdifferential operator theory that we find most satisfactory in terms of obtaining a self-consistent, pure, and elegant definition of discrete scaling that is fully consistent with the theory of the discrete Fourier transform.
\end{abstract}

\section{Introduction}

Signal scaling is a fundamental operation in which the independent variable of the function $f(u)$ is scaled by a real number $M$, resulting in the signal to be compressed or decompressed along the $u$ axis in the form $f(u/M)$. With reference to images, the terms magnification/demagnification or zooming in/out are more commonly used. Scaling or magnification is not trivial for signals of a discrete (integer) variable since, the signal values may not fall onto the discrete coordinate points. Given a function $f[n]$ defined on the integers, the value of $f[n/M]$ will be undefined unless $n/M$ is an integer. Nevertheless, discrete scaling is a necessary operation in practice since we often want to scale signals of continuous variables which are represented as functions of discrete variables in digital computers.

A straightforward approach requiring knowledge of no more than elementary signals and systems \cite{oppenheim96BookSignalsSystems,oppenheim75bookdsp,oppenheim13bookdtsp} is to consider the values of $f[n]$ as the Nyquist-rate samples of a hypothetical bandlimited signal $f(u)$. Then, we can use standard sinc interpolation to write an expression for $f(u)$ in terms of $f[n]$. Now, $f(u)$ can be scaled to $f(u/M)$ and then re-sampled to obtain the values of a new signal of a discrete variable, which can be considered the scaled version of $f[n]$. The values of the new scaled signal will be linearly related to the values of the original signal $f[n]$. Of course, scaling $f(u)$ to obtain $f(u/M)$ will change its bandwidth, which introduces complications in choosing the re-sampling rate. In any case, if the re-sampling rate is different, this will somewhat complicate interpretation of the scaled signal. If the integer domain is not defined from $-\infty$ to $\infty$, but rather over a finite interval, say from $0$ to $N-1$, and we are working in a circulant domain, it is possible to modify the approach by employing Dirichlet functions \cite{ozaktas01book} instead of sinc functions as the interpolation functions.

The only more creative approach to discrete scaling we are aware of is due to Pei et al., who developed a method based on ``Centered Discrete Dilated Hermite Functions'' (CDDHFs) \cite{pei12scaling}, which is an improvement of their earlier ``$n^2$ matrix'' method \cite{pei08commutingmatrices}. The CDDHF-based discrete scaling method works as follows: First, write the signal as a linear superposition of discrete Hermite-Gaussian functions. Then, replace the discrete Hermite-Gaussian functions with their dilated (scaled) versions to obtain the scaled discrete signal \cite{pei12scaling}. In other words, the expansion coefficients are kept the same while scaling the discrete functions that form the expansion basis. Although this sounds conceptually simple, the difficulty (and ingenuity) lies in the development of the the set of dilated discrete Hermite-Gaussian functions, \cite{pei12scaling,mugler02dilatedHFs}, on which the method rests. This procedure provides a mathematically sound and elegant way of performing discrete signal scaling.

In this work, we present a different approach by utilizing hyperdifferential operator theory \cite{ozaktas01book,wolf79book,ozaktas10PoularikasBook,yosida84operationalcalc,nazarathy80operatoralgebra} to obtain a discrete scaling matrix. The scaled version of the signal is obtained by multiplying the unscaled version by this matrix. We choose to work in a framework that is not only discrete, but also finite. That is, the functions are defined over finite intervals. Our approach employs the basic operations of differentiation and coordinate multiplication. We believe that it provides a self-consistent, pure, and elegant definition of discrete scaling which is also fully compatible with the theory of the discrete Fourier transform and its circulant structure. We also believe that the presented approach of defining a discrete operation in the context of hyperdifferential operator theory can set an example that can be applied to other problems in signal theory and analysis.

The paper is organized as follows: preliminaries are given in Section~\ref{prelim}; then in Section~\ref{peimethod} we review Pei's method. Our method is presented in Section~\ref{ourmethod}. Numerical results and comparisons are given in Section~\ref{results}. Finally we conclude in Section~\ref{conclusions}.

\section{Preliminaries}
\label{prelim}

For simplicity we work with one-dimensional signals, although our results can easily be generalized to higher-dimensional signals. Scaling is defined as that operation which takes $f(u)$ to $|M|^{-1/2}f(u/M)$. The factor $|M|^{-1/2}$ is included to make the operation unitary, but this will not be of much importance. The real parameter $M>0$ can be called the scaling or zooming factor or the magnification, depending on context. The signal will be compressed/demagnified or decompressed/magnified depending on whether $M$ is lesser or greater than unity. In operator form we will write
\begin{equation}
{\cal M}_M f(u) = |M|^{-1/2}f(u/M).
\end{equation}
where the calligraphic operator on the left-hand side exhibits the parameter $M$. Our convention for the Fourier transform operator will be
\begin{equation}
\mathcal{F} f(u) = \int_{-\infty}^{\infty} f(u) e^{-j2\pi u \mu}du 
\end{equation}
We define two further operators, the coordinate multiplication operator ${\cal U}$ and the differentiation operator ${\cal D}$:
\begin{eqnarray}
{\cal U} f(u) = uf(u) \\
{\cal D} f(u) = \frac{1}{i2\pi} \frac{df(u)}{du},
\end{eqnarray}
where the $(i2\pi)^{-1}$ is included so that ${\cal U}$ and ${\cal D}$ are precisely Fourier duals (the effect of either in one domain is its dual in the other domain). This duality can be expressed as follows:
\begin{equation} \label{U_fdf}
\mathcal{U}=\mathcal{FDF}^{-1}.
\end{equation}
Basically, the above equation says that, instead of multiplying a function $f(u)$ with $u$, we can instead take its inverse Fourier transform, differentiate it with respect to the frequency variable, divide by $i2\pi$, and take its Fourier transform, and we will get the same result.

In this paper we deal with finite-length signals of a discrete (integer) variable. (We could equivalently think of them as being defined on a circulant domain, which would not make a difference to our arguments.) The length of our signal vectors will be denoted by $N$. When $N$ is even, they will be defined on the interval of integers $[-\frac{N}{2},\frac{N}{2}-1]$, and when $N$ is odd, they will be defined on the interval of integers $[-\frac{N-1}{2},\frac{N-1}{2}]$. We will also consider a less-common approach based on the device of using ``half integers.'' In this approach, the domain is defined as the interval of unit-spaced half integers $[-\frac{N}{2}+0.5,\frac{N}{2}-1+0.5]$ for even $N$ and $[-\frac{N-1}{2}-0.5,\frac{N-1}{2}-0.5]$ for odd $N$. Although not very usual, there is nothing unnatural about this way of indexing signals of a discrete variable; it is merely a particular way of bookkeeping. Note that the indices are still spaced by unity, and there is merely a shift by $0.5$ with the purpose of making the interval symmetrical around the origin when $N$ is even (with the consequence that symmetry is lost when $N$ is odd). A few examples of works considering this way of indexing are \cite{grunbaum82centeredDFT,clary03shiftedFourier,mugler11centeredDFT,vargasrubio05centered}. Consistent with this literature, we will refer to the former approach as the \textit{ordinary\/} DFT and refer to the latter one, in which we use "half integers", as the \textit{centered\/} DFT.

It is possible to better understand the choices of indexing by considering them in the context of sampling a signal of a continuous (real) variable. The sample values of a function $f(u)$ are usually written as $f(nh)$ where $n$ is the index and $h$ is the sampling interval. When we use full integer values of the index, which is the usual case, we get a set of samples that includes a sample at the origin, $f(0)$ for $n=0$. For instance, for $N=4$, we would be sampling at $u=-2h,-1h,0h,1h$. However, we may also choose to sample in a manner that does not include the origin, for instance, we may choose our samples as $f(nh+0.5)$, where $n$ are still full integers, in which case we would be sampling at $u=-1.5h,-0.5h,0.5h,1.5h$. The use of half integers is an alternative way of bookkeeping where we maintain the samples to be at $f(nh)$ rather than $f(nh+0.5)$, but still get the same samples by allowing $n$ to take half integer values. While both approaches are equivalent, we find the use of half integers (centered) to be more elegant and unifying.

The sampled signals can be represented by column vectors with $N$ rows. The labelling of the rows will follow the same index convention as above. In the case of half integers, we may refer to the ``-1.5th row'' of the vector, and so forth. The operators acting on them can be represented as matrices that have $N$ columns and $N$ rows. The matrix representing the Fourier transformation will be the unitary discrete Fourier transform (DFT) matrix $\mathbf{F}$, with appropriate shifting/circulation of its rows and columns such that it is consistent with the index ranges we use. The elements $F_{mn}$ of this $N$-point unitary DFT matrix $\mathbf{F}$ can be written in terms of $W_N = \exp(-j 2\pi/N)$ as follows:
\begin{equation*} \label{DFT}
F_{mn}= \frac{1}{\sqrt{N}\,} W_N^{mn}.
\end{equation*}
Note that this expression covers both the ordinary and centered case provided we remember that (i) for the ordinary case $m$ and $n$ run through the integers $[-\frac{N}{2},\frac{N}{2}-1]$ for even $N$ and $[-\frac{N-1}{2},\frac{N-1}{2}]$ for odd $N$; (ii) for the centered case $m$ and $n$ run through the unit-space half integers $[-\frac{N}{2}+0.5,\frac{N}{2}-1+0.5]$ for even $N$ and $[-\frac{N-1}{2}-0.5,\frac{N-1}{2}-0.5]$ for odd $N$. The ability to write what would otherwise be two separate expressions in the familiar form above is the main advantage of the half-integer indexing scheme we employ.

It could be questioned whether it would not be simpler to work with the traditional interval $[0,N-1]$ to keep things simple. This would essentially give the same results, only in shifted/circulated form. We choose to work with symmetric intervals to maintain and reveal as much symmetry in the problem as there actually is.

We work with dimensionless coordinates; that is, the unit of $u$ is not seconds or meters, it is unitless. Say the function $\hat{f}(x)$ of a continuous variable $x$ in seconds or meters has an approximate extent lying over the interval $[-\Delta x/2,\Delta x/2]$, meaning most of its energy is contained in this interval. Likewise, say its extent in the frequency domain lies over the interval $[-\Delta f/2,\Delta f/2]$, where $f$ is the frequency variable in Hz or inverse meters. Then we can introduce a parameter $s$, such that $u=x/s$ is a dimensionless number and choose to work with the function $f(u) = \hat{f}(su)$ instead of $\hat{f}(x)$. If we choose $s=\sqrt{\Delta x/\Delta f}\,$, then the extent of {\em both\/} $f(u)$ and its Fourier transform will lie in the interval $[-\sqrt{\Delta x\Delta f}\,/2,\sqrt{\Delta x\Delta f}\,/2]$. According to the sampling theorem, if a signal is contained within such an interval, it can be sampled with a sampling interval of $1/\sqrt{\Delta x\Delta f}\,$. Thus there will be $N= \sqrt{\Delta x\Delta f}\,/(1/\sqrt{\Delta x\Delta f}\,) = \Delta x\Delta f$ samples in all. The quantity $\Delta x\Delta f$ is often referred to as the time-bandwidth or space-bandwidth product. Re-expressing in terms of the number of samples $N$, we would be sampling over the interval $[-\sqrt{N}\,/2,\sqrt{N}\,/2]$ with a sampling interval of $1/\sqrt{N}\,$ for a total of $N$ samples. Should $N$ not have an even whole square root, we can always choose $\Delta x$ and $\Delta f$ a little larger than necessary to make it so, although this will not be important for our discussion.

To put the whole sampling issue together, let us consider the example of $N=16$. The interval over which the signal will be sampled will be $[-2,2]$, which is divided into 16 sampling intervals each of length $h=1/4$. The real issue now is whether the samples will be taken on the left (or right) edge of each sampling interval, or in the center (or yet somewhere else) of each sampling interval. Taking them at the left edge is the familiar case; the sample points will be $[-2,-1.75, \ldots, 0, \ldots, 1.5, 1.75]$. If we take them at the middle, they will be $[-1.875, -1.625, \ldots, -0.125, 0.125, \ldots, 1.875]$. There are two ways to bookkeep the latter case. We can continue to work with an integer index $n$, and then the sample points will be $u=(n+1/2)h$. Alternatively, we can maintain that the sample points are still at $u=nh$, but use half integer values of $n$. We find greater clarity and unity in emphasizing the sampling intervals over the sampling points, and working with half integer index values.

\section{Pei's Method}
\label{peimethod}
In \cite{pei12scaling}, Pei et al. consider a finite signal denoted $f$, of length $N$, to be scaled. They let $f_M$ denote the scaled signal, with $M$ being the scaling factor. The signal $f$ can always be expressed as a linear combination of any $N$ linearly independent signals. In their method, special functions called ``Centered Discrete Dilated Hermite Functions'' (CDDHFs) are utilized as the set of $N$ linearly independent signals so any $f$ can be expressed as a linear combination of CDDHFs:
\begin{equation}\label{x_expansion}
f = \sum_{p=0}^{N-1} c_{p,1}H_{p,1} \quad,
\end{equation}
where the $H_{p,1}$ are the CDDHFs of length $N$, and the $c_{p,1}$ are the coefficients. The coefficients are simply the inner products of the $H_{p,1}$ with the signal $x$: that is, $c_{p,1}=\langle x, H_{p,1} \rangle$. The CDDHFs $H_{p,1}$ are the centered discrete Hermite functions with no scaling; hence the subscript $1$ in the notation. $H_{p,M}$ denotes the $p$th CDDHF with a scaling factor of $M$.

Their proposed way to scale the signal $f$ is to scale the basis signals $H_{p,1}$, and keep the expansion coefficients the same. More explicitly, the scaled signal $f_M$ is obtained as follows:
\begin{equation}
f_M = \sum_{p=0}^{N-1} c_{p,1}H_{p,M} \quad,
\end{equation}
where the $H_{p,M}$ are scaled versions of the $H_{p,1}$ with a scaling factor of $M$. The critical task, of course, is to find the scaled versions of the basis signals. Pei et al. develop a method for constructing the CDDHFs $H_{p,M}$, which we now summarize.

The Hermite-Gaussian functions of a continuous variable, denoted by $\psi_p(u)$, are given as follows, \cite{ozaktas01book}:
\begin{equation}
	\psi_p(u) = A_p H_p(\sqrt{2\pi}u)e^{-\pi u^2}, \hspace{3mm} A_p = \dfrac{2^{1/4}}{\sqrt{2^p p!}}
\end{equation}
where $H_p(u)$ denotes the Hermite polynomials.
It is well-known that the Hermite-Gaussian functions, $\psi_p(u)$, satisfy the differential equation \cite{ozaktas01book}
\begin{equation} \label{diff_eq1}
\frac{d^2}{du^2}\psi_p(u) - 4\pi^2u^2\psi_p(u) = \lambda \psi_p(u).
\end{equation}
The time-scaled version of the Hermite-Gaussian function $\psi_p(u)$ is $\psi_p(u/M)$. It is possible to find a differential equation for $\psi_p(u/M)$ by simply replacing every $u$ by $u/M$ in Eq.~(\ref{diff_eq1}):
\begin{equation} \label{diff_eq2}
M^2\frac{d^2}{du^2}\psi_p\left(\frac{u}{M}\right) -4\pi^2 \left(\frac{u}{M}\right)^2\psi_p\left(\frac{u}{M}\right) = \lambda \psi_p\left(\frac{u}{M}\right).
\end{equation}
Eq. (\ref{diff_eq2}) can be rewritten in terms of the coordinate multiplication operator $\mathcal{U}$ and the differentiation operator $\mathcal{D}$:
\begin{equation}
(-M^24\pi^2\mathcal{D}^2-\frac{4\pi^2}{M^2}\mathcal{U}^2) \psi_p\left(\frac{u}{M}\right) = \lambda \psi_p\left(\frac{u}{M}\right).
\end{equation}
Rearranging the terms, we get:
\begin{equation} \label{oper_eq}
(M^4\mathcal{D}^2+\mathcal{U}^2) \psi_p\left(\frac{u}{M}\right) = -\frac{M^2}{4\pi^2}\lambda \psi_p\left(\frac{u}{M}\right).
\end{equation}
Functions $\psi_p(t/M)$ satisfying Eq. (\ref{oper_eq}) are the eigenfunctions of $(M^4\mathcal{D}^2+\mathcal{U}^2)$.

The next step is to find the discrete counterpart of Eq.~(\ref{oper_eq}). This is done by replacing the abstract operators (denoted by calligraphic letters), by boldface matrix operators that act on column vectors in the form $(M^4\mathbf{D}^2+\mathbf{U}^2)$. Then, it is possible to compute the CDDHFs $H_{p,M}$ as the eigenvectors of this matrix. Here $\mathbf{U}$ and $\mathbf{D}$ are matrices that are the finite discrete manifestations of the abstract operators $\mathcal{U}$ and $\mathcal{D}$. So the remaining task before implementing the method is to determine what $\mathbf{U}$ and $\mathbf{D}$ should be.

Pei et al. define the matrix $\mathbf{U}^2$ as follows:
\begin{equation}\label{peiu2}
 \mathbf{U}^2_{mn} = 
\begin{cases} 
      \left(m-\frac{N-1}{2}\right)^2 & \text{if } m=n \\
      0 & \text{otherwise}, \\
   \end{cases}
\end{equation}
where $\mathbf{U}^2_{mn}$ is the $m$th row, $n$th column entry of $\mathbf{U}^2$, and $m,n = 0, 1, \ldots\ N-1$.
Intuitively, this corresponds to multiplying every entry in a signal by the square of the corresponding index in a centered manner (hence the $-(N-1)/2$ term). (It will be interesting to contrast this with our development of the $\mathbf{U}$ matrix later on. We do not take for granted that $\mathbf{U}$ should be a simple reflection of the form of the continuous manifestation of the $\mathcal{U}$ operator, and indeed show that for a formulation satisfying complete structural symmetry, it should be chosen differently.)

Once $\mathbf{U}^2$ is defined, we have $\mathbf{D}^2= \mathbf{F}\mathbf{U}^2\mathbf{F}^{-1}$ by using the duality relation given in Eq.~\ref{U_fdf}. Being the finite discrete manifestation of the abstract operator $\mathcal{F}$, the matrix $\mathbf{F}$ is the standard centered DFT matrix. Finally, for any scaling factor $M$, we can form $(M^4\mathbf{D}^2+\mathbf{U}^2)$, and find its eigenvectors $H_{p,M}$, after which we can easily complete the process. More on the implementation details of this approach can be found in~\cite{pei12scaling}.

\section{Hyperdifferential Operator Based Matrix Method}
\label{ourmethod}

It is an established fact that the scaling operator $\mathcal{M}_M$ can be written in hyperdifferential form as follows in terms of the $\mathcal{U}$ and $\mathcal{D}$ operators \cite{wolf79book,ozaktas01book,yosida84operationalcalc,nazarathy80operatoralgebra}:
\begin{equation} \label{scaling_op}
\mathcal{M}_M=\exp{\left(-i2\pi \ln{(M)} \,
\frac{\mathcal{UD}+\mathcal{DU}}{2}\right)}.
\end{equation}

Our approach is based on requiring that all the discrete entities we define observe the same operational properties and relationships as they do in abstract operator form. Therefore, we will require the discrete manifestations of Eq.~(\ref{U_fdf}) and Eq.~(\ref{scaling_op}) to have the same structure, with the abstract operators being replaced by matrix operators. As a consequence, Eq.~(\ref{U_fdf}) will hold for finite difference and matrix versions of the ${\cal D}$ and ${\cal U}$ operators and the matrix operator counterpart of ${\cal M}_M$ will be
\begin{equation} \label{scaling_mat}
\mathbf{M}_M=\exp{\left(-i2\pi \ln{(M)} \,
\frac{\mathbf{UD}+\mathbf{DU}}{2}\right)}.
\end{equation}
Thus, to scale a function of a discrete variable, we need to write it as a column vector and multiply it with the scaling matrix $\mathbf{M}_M$. In order to obtain the scaling matrix, we need the first-order differentiation and coordinate multiplication matrices $\mathbf{D}$ and $\mathbf{U}$ and then compute the matrix exponential of the expression inside the parentheses. Therefore our first task is to obtain the $\mathbf{D}$ and $\mathbf{U}$ matrices.

For signals of discrete variables, the closest thing to differentiation is finite differencing. Consider the following definition:
\begin{equation} \label{D_defn}
\tilde{\mathcal{D}}_hf(u)=\frac{1}{i2\pi}\frac{f(u+h/2)-f(u-h/2)}{h}.
\end{equation}
If $h\rightarrow 0$, then $\tilde{\mathcal{D}}_h \rightarrow \mathcal{D}$, since in this case the right-hand side approaches $(i2\pi)^{-1}df(u)/du$. Therefore, $\tilde{\mathcal{D}}_h$ can be interpreted as a finite difference operator.

Now, using $f(u+h)=\exp(i2\pi h\mathcal{D})f(u)$, which is another established result in operator theory \cite{wolf79book,ozaktas01book}, we express Eq.~(\ref{D_defn}) in hyperdifferential form:
\begin{align} \label{D_hyp}
\tilde{\mathcal{D}}_h & =\frac{1}{i2\pi}
\frac{e^{i\pi h\mathcal{D}}-e^{-i\pi h\mathcal{D}}}{h} \nonumber \\
& = \frac{1}{i2\pi} \frac{2i\sin(\pi h\mathcal{D})}{h}
={\rm sinc}(h\mathcal{D}) \;\mathcal{D}.
\end{align}
Note that if we let $h\rightarrow 0$ in the last equation and take the limit, we can verify that $\tilde{\mathcal{D}}_h \rightarrow \mathcal{D}$ from here as well. 

Now, we turn our attention to the task of defining $\tilde{\mathcal{U}}_h$. It is tempting to define the discrete version of the coordinate multiplication matrix by simply forming a diagonal matrix with the diagonal entries being equal to the coordinate values, with due adjustment for centering and discreteness, much as in Eq.~(\ref{peiu2}). However, upon closer inspection we have decided that this could not be taken for granted. In order to obtain the most elegant and purest formulation possible, we must be sure to maintain the structural symmetry between $\mathcal{U}$ and $\mathcal{D}$ in all their manifestations. Therefore, we choose to define $\tilde{\mathcal{U}}_h$ such that it is related to $\mathcal{U}$, in exactly the same way as $\tilde{\mathcal{D}}_h$ is related to $\mathcal{D}$:
\begin{equation}
\tilde{\mathcal{U}}_h = {\rm sinc}(h\mathcal{U}) \;\mathcal{U},
\end{equation}
from which we can observe that as $h\rightarrow 0$, we have $\tilde{\mathcal{U}}_h \rightarrow \mathcal{U}$, as should be. But beyond that, it is also possible to show that, $\tilde{\mathcal{U}}_h$, when defined like this, satisfies the same duality expression Eq.~(\ref{U_fdf}) satisfied by $\mathcal{U}$ and $\mathcal{D}$:
\begin{equation}
\tilde{\mathcal{U}}_h = \mathcal{F} \tilde{\mathcal{D}}_h \mathcal{F}^{-1}
\end{equation}
To see this, substitute $\tilde{\mathcal{D}}_h$ in this equation:
\begin{align}
\tilde{\mathcal{U}}_h & = \mathcal{F} \left(\frac{1}{i2\pi} \frac{2i\sin(\pi h\mathcal{D})}{h}\right)  \mathcal{F}^{-1} \nonumber \\
& = \frac{1}{i2\pi} \frac{2i\sin(\pi h\mathcal{U})}{h}
={\rm sinc}(h\mathcal{U}) \mathcal{U}.
\end{align}
When acting on a continuous signal $f(u)$, the operator $\mathcal{U}$ becomes
\begin{equation} \label{U_final_op}
\tilde{\mathcal{U}}_hf(u)=\frac{1}{\pi} \frac{\sin(\pi hu)}{h} f(u).
\end{equation}
We observe the effect is not merely multiplying with the coordinate variable. Had we defined $\tilde{\mathcal{U}}_h$ such that it corresponds to multiplication with the coordinate variable, we would have destroyed the symmetry and duality between $\mathcal{U}$ and $\mathcal{D}$ in passing to the discrete world.

Now, by sampling Eq.~(\ref{U_final_op}), we can obtain the matrix operator to act on finite discrete signals. The sample points will be taken as $u=nh$ with the range of $n$ being determined by whether the number of sample $N$ is even or odd, and by whether we use the ordinary or centered sampling scheme, as explained in detail in Section~\ref{prelim}. Finally, we are able to write the elements of the matrix $\mathbf{U}$:
\begin{equation}
U_{mn}= \begin{cases}
\frac{\sqrt{N}\,\;}{\pi} \sin \left(\frac{\pi}{N}n \right), & \text{for } m=n \\
0, & \text{for } m \neq n
\end{cases}.
\end{equation}

The next step is to obtain the $\mathbf{D}$ matrix. To do so, first recall that Eq.~\ref{U_fdf} can also be written as 
\begin{equation} \label{D_fuf}
\mathcal{D}=\mathcal{F}^{-1}\mathcal{UF}.
\end{equation}
Since we want the finite discrete manifestations of these abstract operators to also exhibit the same structure, we write
\begin{equation} \label{D_fuf_matrix}
\mathbf{D}=\mathbf{F}^{-1}\mathbf{UF},
\end{equation}
where $\mathbf{F}$ was defined in Eq.~(\ref{DFT}). Thus, we have now obtained discrete matrix forms $\mathbf{U}$ and $\mathbf{D}$ of the coordinate multiplication and differentiation operators so we are finally in a position to calculate the discrete scaling operator defined in Eq.~(\ref{scaling_mat}).

Before we move on to numerical results and interpretations, several comments will be in order. First of all, it will be worth recapitulating what we did and why. As mentioned, it is tempting to define the discrete version of the coordinate multiplication matrix by simply forming a diagonal matrix with the diagonal entries being equal to the coordinate values. Then one could also have easily obtained the discrete version of the differentiation matrix by using duality, without having to go through the circuitous route we followed. However, due to the circulant structure of the finite/periodic lattice associated with the DFT, we suspected this may not be true and decided to begin with the differentiation matrix instead. The simplest way to define the finite difference operator would be, instead of Eq.~(\ref{D_defn}),
\begin{equation}
\tilde{\mathcal{D}}_hf(u)=\frac{1}{i2\pi}\frac{f(u+h)-f(u)}{h}.
\end{equation}
However, when discretized, the corresponding differentiation matrix would have values of $-1$ along the primary diagonal and values of $1$ along the diagonal adjacent to the primary, leaving us with a matrix that is not symmetric. We rejected this option since it would clearly not give us a pure and elegant formulation, opting for Eq.~(\ref{D_defn}) instead. However, this definition, while symmetric, did not allow us to immediately write a differentiation matrix, because it involved sample points in the middle of the sampling intervals, rather than the ends. Fortunately, the relationship Eq.~(\ref{D_hyp}) between $\tilde{\mathcal{D}}_h$ and $\mathcal{D}$ that we derived showed us the way to define $\tilde{\mathcal{U}}_h$. The operator $\tilde{\mathcal{U}}_h$ did not exhibit the same problem of involving sample points in the middle that $\tilde{\mathcal{D}}_h$ did, and could be discretized without difficulty. It was also symmetrical, as we desired it to be. Once we obtained the $\mathbf{U}$ matrix, it was possible to use duality to obtain the $\mathbf{D}$ matrix as well.

We believe that the presented way of defining the finite matrix forms of the coordinate multiplication and differentiation operators is the only way consistent with the circulant structure of the DFT and the dual nature of these operators.

\section{Quantitative Discussion}
\label{results}

In this section, we examine our formulation from a numerical perspective.
We consider two different functions: a chirped pulse function $\exp(-\pi u^2-j \pi u^2)$, denoted by F1, and the trapezoidal function $1.5{\rm tri}(u/2)-0.5{\rm tri}(2u)$, denoted by F2 (${\rm tri}(u)={\rm rect}(u)*{\rm rect}(u)$). We considered three different values for the scale parameter $M$: $0.5, 2, 3$. Analytically-derived scaled versions for our two functions are taken as the comparison reference. We calculated normalized mean-square errors (MSE) between the following vectors: (i) Reference: Samples of the continuously scaled functions $f(u/M)$; (ii) Discrete scaling: The product of the samples of the original function $f(u)$ with the discrete scaling matrix. As explained in Section~\ref{prelim}, results are calculated for both \textit{centered} and \textit{ordinary} sampling regimes. The number of samples $N$ are taken as 128, 256, and 512. Results are tabulated as percentages in Tables~\ref{mse_scores_tableF1} to \ref{mse_scores_tableF2}.

The results confirm that our approach for discrete scaling formulation approximates the continuous scaling reasonably well. If higher accuracies are needed, one can always increase $N$ and make a better approximation as the MSE decreases with increasing $N$. This is because large $N$ means larger extents in both the time and frequency domains, so that a smaller percentage of the signal is left outside of these extents. Moreover, as expected, the MSE values also depend on the signal that is being scaled. Recalling the information theoretic considerations given in Section~\ref{prelim}, the accuracy obtained depends on what percentage of the signal energy is confined within the chosen extents in the time and frequency domains. For example, MSE values for F2 are relatively higher than those of F1. This is caused by the fact that its frequency domain content is spread over a relatively greater extent, leading to a greater percentage of its energy to fall outside the chosen extents. As can be observed, use of either the centered or ordinary approaches gives similar results, as this choice does not make any essential difference with regards to accuracy.

\begin{table}[!t]
\renewcommand{\arraystretch}{1.3}
\centering \caption{Percentage MSE Scores - Chirped Pulse }
\label{mse_scores_tableF1}
\begin{tabular}{llllll}
\hline
 Parameter \textit{M}  & $\:$ $\:$ N  & Centered & Ordinary \\
\hline
\multirow{3}{*}{2} & \vline $\:$ 128 $\:$\vline &$1.22\times10^{-2}$&$1.22\times10^{-2}$\\
                    & \vline $\:$ 256 $\:$\vline &$3.1\times10^{-3}$&$3.1\times10^{-3}$\\
                    & \vline $\:$ 512 $\:$\vline &$7.75\times10^{-4}$&$7.75\times10^{-4}$\\
\hline
\multirow{3}{*}{3} & \vline $\:$ 128 $\:$\vline &$6.36\times10^{-2}$&$6.36\times10^{-2}$\\
                    & \vline $\:$ 256 $\:$\vline &$1.62\times10^{-2}$&$1.62\times10^{-2}$\\
                    & \vline $\:$ 512 $\:$\vline &$4.1\times10^{-3}$&$4.1\times10^{-3}$\\
\hline
\multirow{3}{*}{0.5} & \vline $\:$ 128 $\:$\vline &$2.68\times10^{-2}$&$2.68\times10^{-2}$\\
                    & \vline $\:$ 256 $\:$\vline &$6.9\times10^{-3}$&$6.9\times10^{-3}$\\
                    & \vline $\:$ 512 $\:$\vline &$1.8\times10^{-3}$&$1.8\times10^{-3}$\\
\hline
\end{tabular}
\end{table}

\begin{table}[!t]
\renewcommand{\arraystretch}{1.3}
\centering \caption{Percentage MSE Scores - Trapezoid}
\label{mse_scores_tableF2}
\begin{tabular}{llllll}
\hline
 Parameter \textit{M}  &  $\:$ $\:$ N  & Centered & Ordinary \\
\hline
\multirow{3}{*}{2} & \vline $\:$ 128 $\:$\vline &$0.33$&$0.313$\\
                    & \vline $\:$ 256 $\:$\vline &$9.47\times10^{-2}$&$0.103$\\
                    & \vline $\:$ 512 $\:$\vline &$2.91\times10^{-2}$&$2.92\times10^{-2}$\\
\hline
\multirow{3}{*}{3} & \vline $\:$ 128 $\:$\vline &$1.62$&$1.59$\\
                    & \vline $\:$ 256 $\: $\vline &$0.51$&$0.53$\\
                    & \vline $\:$ 512 $\: $\vline &$0.15$&$0.15$\\
\hline
\multirow{3}{*}{0.5} & \vline $\:$ 128 $\:$\vline &$4.21\times10^{-2}$&$4.75\times10^{-2}$\\
                    & \vline $\:$ 256 $\:$\vline &$1.69\times10^{-2}$&$3.16\times10^{-2}$\\
                    & \vline $\:$ 512 $\:$\vline &$1.2\times10^{-2}$&$7.4\times10^{-3}$\\

\hline
\end{tabular}
\end{table}

\section{Conclusion}
\label{conclusions}

In this paper, a formulation for scaling of discrete-time signals based on hyperdifferential operator theory is proposed. For finite-length signals of a discrete variable, a unitary scaling matrix is obtained so that the scaled version can be obtained by a direct matrix multiplication. Given the vector holding the samples of the unscaled signal, this scaling matrix  multiplies the input vector to obtain the samples of the scaled signal. We also discussed two different approaches to indexing the discrete signals, namely \textit{ordinary} indexing and \textit{centered} indexing. These indexing approaches are fully consistent with the well-known ordinary and centered discrete Fourier transform (DFT) definitions. Furthermore, the proposed formulation is mathematically elegant, pure and uses self-consistent coordinate multiplication and differentiation operations. If needed, depending on the application, the accuracy of the resulting method can be improved by using coordinate multiplication and differentiation matrices that are obtained by brute force numerical approximations to the continuous domain. However, in this paper our purpose was to demonstrate these matrices in their purest forms without any numerical approximation.

We believe that we have obtained an elegant and pure formulation of discrete scaling based on self-consistent definitions of coordinate multiplication and differentiation operators. Our approach is consistent with the circulant nature of the discrete Fourier transform and also provides numerically satisfactory results.

\bibliographystyle{plain}
\bibliography{archives}

\begin{thebibliography}{10}

\bibitem{clary03shiftedFourier}
Stuart Clary and Dale~H. Mugler.
\newblock Shifted fourier matrices and their tridiagonal commutors.
\newblock {\em SIAM Journal on Matrix Analysis and Applications},
  24(3):809--821, 2003.

\bibitem{grunbaum82centeredDFT}
F.Alberto Grunbaum.
\newblock The eigenvectors of the discrete fourier transform: A version of the
  hermite functions.
\newblock {\em Journal of Mathematical Analysis and Applications}, 88(2):355 --
  363, 1982.

\bibitem{mugler11centeredDFT}
D.~H. Mugler.
\newblock The centered discrete fourier transform and a parallel implementation
  of the fft.
\newblock In {\em 2011 IEEE International Conference on Acoustics, Speech and
  Signal Processing (ICASSP)}, pages 1725--1728, May 2011.

\bibitem{mugler02dilatedHFs}
D.~H. Mugler, S.~Clary, and Y.~Wu.
\newblock Discrete hermite expansion of digital signals: Applications to ecg
  signals.
\newblock In {\em IEEE Signal Process. Soc. 10th DSP Workshop}, pages 262--267,
  2002.

\bibitem{nazarathy80operatoralgebra}
M.~Nazarathy and J.~Shamir.
\newblock Fourier optics described by operator algebra.
\newblock {\em J. Opt. Soc. Am.}, 70:150--159, 1980.

\bibitem{oppenheim75bookdsp}
Alan~V. Oppenheim and Ronald~W. Schafer.
\newblock {\em Digital Signal Processing}.
\newblock Prentice-Hall, Inc., 1975.

\bibitem{oppenheim13bookdtsp}
Alan~V. Oppenheim and Ronald~W. Schafer.
\newblock {\em Discrete-time Signal Processing (3rd Ed.)}.
\newblock Pearson Education Limited, 2013.

\bibitem{oppenheim96BookSignalsSystems}
Alan~V. Oppenheim, Alan~S. Willsky, and S.~Hamid Nawab.
\newblock {\em Signals \& Systems (2nd Ed.)}.
\newblock Prentice-Hall, Inc., Upper Saddle River, NJ, USA, 1996.

\bibitem{ozaktas01book}
H.~M. Ozaktas, Z.~Zalevsky, and M.~A. Kutay.
\newblock {\em The Fractional Fourier Transform with Applications in Optics and
  Signal Processing}.
\newblock New York: Wiley, 2001.

\bibitem{ozaktas10PoularikasBook}
Haldun~M. Ozaktas, M.~Alper Kutay, and Cagatay Candan.
\newblock {\em Transforms and Applications Handbook}, chapter Fractional
  Fourier Transform, pages 14--1--14--28.
\newblock CRC Press, Boca Raton, New York, NY, 2010.

\bibitem{pei08commutingmatrices}
S.~C. Pei, J.~J. Ding, W.~L. Hsue, and K.~W. Chang.
\newblock Generalized commuting matrices and their eigenvectors for dfts,
  offset dfts, and other periodic operations.
\newblock {\em Signal Processing, IEEE Transactions on}, 56(8):3891 -- 3904,
  2008.

\bibitem{pei12scaling}
S.~C. Pei and Y.~Lai.
\newblock Signal scaling by centered discrete dilated hermite functions.
\newblock {\em IEEE Trans. Signal Process.}, 60:498--503, 2012.

\bibitem{vargasrubio05centered}
J.~G. Vargas-Rubio and B.~Santhanam.
\newblock On the multiangle centered discrete fractional fourier transform.
\newblock {\em IEEE Signal Process. Lett.}, 12(4):273--276, 2005.

\bibitem{wolf79book}
K.~B. Wolf.
\newblock {\em Integral Transforms in Science and Engineering (Chapter~9:
  Construction and properties of canonical transforms)}.
\newblock New York: Plenum Press, 1979.

\bibitem{yosida84operationalcalc}
K.~Yosida.
\newblock {\em Operational Calculus: A Theory of Hyperfunctions}.
\newblock Springer, New York, USA, 1984.

\end{thebibliography}

\end{document}